\documentclass{article}
\usepackage[affil-it]{authblk}

\providecommand{\keywords}[1]
{
  \small	
  \textbf{\textit{Keywords:}} #1
}

\usepackage{epsfig}
\usepackage{subcaption}
\usepackage{calc}
\usepackage{amssymb}
\usepackage{amstext}
\usepackage{amsmath}
\usepackage{amsthm}
\usepackage{multicol}
\usepackage{pslatex}
\usepackage{apalike}
\usepackage{float}

\usepackage{comment}

\usepackage{hyperref}

\usepackage{epsfig}
\usepackage{url}
\usepackage{xspace,xargs}
\usepackage{todonotes}
\usepackage{blindtext}
\usepackage[normalem]{ulem}

\usepackage{multirow}

\usepackage{caption}

\usepackage{booktabs}

\def\BWT{\textrm{BWT}\xspace}
\def\EBWT{\textrm{EBWT}\xspace}
\def\LCP{\textrm{LCP}\xspace}

\def\bigS{\mathcal{S}\xspace}

\def\ebwt{{\normalfont\textsf{ebwt}}\xspace}
\def\lcp{{\normalfont\textsf{lcp}}\xspace}

\def\qs{{\normalfont\textsf{qs}}\xspace}

\def\fqs{\textrm{FQSquezeer}\xspace}

\def\fastQC{\textsc{BFQzip}\xspace}

\def\LCPVmin{\textrm{B$_{min}$}}
\def\LCPVthr{\textrm{B$_{thr}$}}

\def\PREC{\textrm{PREC}\xspace}
\def\F1{\textrm{F}\xspace}
\def\SEN{\textrm{SEN}\xspace}

\newcommand{\ie}{{\it i.e.}\xspace}
\newcommand{\eg}{{\it e.g.}\xspace}

\begin{document}

\title{Lossy Compressor preserving variant calling through Extended BWT}

\date{}

\author[1]{Veronica Guerrini\thanks{
\url{veronica.guerrini@di.unipi.it}; Corresponding author
}}
\affil[1]{Department of Computer Science,  University of Pisa, Italy}

\author[2]{Felipe A. Louza}
\affil[2]{Federal University of Uberl\^andia, Brazil}

\author[1]{Giovanna Rosone}


\maketitle


\abstract{
\noindent
A standard format used for storing the output of high-throughput sequencing experiments is the FASTQ format. It comprises three main components: (i) {\it headers}, (ii)  {\it bases} (nucleotide sequences), and (iii) {\it quality scores}.
FASTQ files are widely used for variant calling, where sequencing data are mapped into a reference genome to discover variants that may be used for further analysis.
There are many specialized compressors that exploit redundancy in FASTQ data with the focus only on either the {\it bases} or the {\it quality scores} components.
In this paper we consider the novel problem of lossy compressing, in a reference-free way, FASTQ data by modifying both components at the same time, while preserving the important information of the original FASTQ.
We introduce a general strategy, based on the Extended Burrows-Wheeler Transform (EBWT) and positional clustering, and we present implementations in both internal memory and external memory.
Experimental results show that the lossy compression performed by our tool is able to achieve good compression while preserving information relating to variant calling more than the competitors.
\newline
\newline
{\bf Availability:} the software is freely available at \url{https://github.com/veronicaguerrini/BFQzip}.
}

\section{Introduction}

The recent improvements in high-throughput sequencing technologies have led a reduced cost of DNA sequencing and unprecedented amounts of genomic datasets, which has motivated the development of new strategies and tools for compressing these data that achieve better results than general-purpose compression tools -- see \cite{surveyCompr2016,Hernaez2019} for good reviews.

FASTQ is the standard text-based format used to store raw sequencing data, each DNA fragment ({\it read}) is stored in a record composed by three main components: 
(i) read identifier with information related to the sequencing process ({\it header}), 
(ii) nucleotide sequence ({\it bases}), and 
(iii) quality sequence, with a per-base estimation of sequencing confidence ({\it quality scores}). 
The last two components are divided by a ``separator'' line, which is generally discarded by compressors as it contains only a ``+'' symbol optionally followed by the same header. 

The majority of compressors for FASTQ files commonly split the data into those three main components (or {\it streams}), and compress them separately, which allows much better compression rates.

The {\it headers} can be efficiently compressed taking advantage of their structure and high redundancy.
A common strategy used by FASTQ compressors, like SPRING \cite{SPRING2018} and FaStore \cite{FaStore2018}, is to tokenize each header: the separators are non-alphanumerical symbols \cite{Bonfield2013}.   

The {\it bases} and {\it quality scores} are commonly processed separately, although their information are correlated,
and current specialized compressors only focus on one of these two components.

Most of FASTQ compressors limit their focus on the {\it bases}, compressing the {\it quality scores} independently with a third tool or standard straightforward techniques. 
The approaches that focus on compression the {\it bases} component are {\it lossless}, \ie, they do not modify the bases, but they find a good strategy to represent the data by exploiting the redundancy of the DNA sequences.
An interesting strategy is to reorder the sequences in the 
FASTQ file to gather reads originating from close regions of the genome 
\cite{CoxBJR12,FaStore2018,SCALCE2012,SPRING2018}. 

On the other hand, approaches that focus on compression the {\it quality score} component are generally {\it lossy}, \ie, they modify the data by smoothing the quality scores whenever possible.
These approaches can be {\it reference-based} when they use external information (besides the FASTQ itself), such as a reference corpus of $k$-mers,
\eg QUARTZ~\cite{QUARTZ2015}, GeneCodeq \cite{GeneCodeq2016} and YALFF \cite{YALFF2019}.
While, {\it  reference-free} strategies evaluate only the {\it quality scores} information, such as the quantization of quality values using QVZ \cite{QVZ2015}, Illumina 8-level binning 
and binary thresholding, or evaluate the related biological information in the {\it bases} component (strategies known as {\it  read-based}) -- \eg BEETL \cite{JaninRosoneCox2014} and LEON \cite{LEON2015}.

\paragraph{\bf Our contribution.} 
In this work, we focus on a novel approach for the lossy compression of both the {\it bases} and the {\it quality scores} components taking into account both information at the same time.
In fact, the two components are highly correlated being the second one a confidence estimation of each base call contained in the first one.
To the best of our knowledge, none of the existing FASTQ compressor tools evaluates and modifies both components at the same time.

Note that we are not interested in compressing the {\it headers} component, for which one can use any state-of-the-art strategies or ignore them. Indeed, headers can be artificially structured in fields to store information only related to the sequencing process.

We focus on lossy reference-free and read-based FASTQ compression that makes clever modifications on the data, by reducing noise in the {\it bases} component that could be introduced by the sequencer, and by smoothing irrelevant values on the {\it quality scores} component, according to the correlated information in the {\it bases}, that would guarantee to preserve variant calling.

Hence, we propose a novel read-based, reference- and assembly-free compression approach for FASTQ files, \fastQC, which combines both DNA bases and quality information for obtaining a lossy compressor.

Similarly to BEETL \cite{JaninRosoneCox2014}, our approach is based on 
the Extended Burrows-Wheeler Transform (EBWT)~\cite{bwt94,MantaciRRS07}
and its combinatorial properties, and it applies the idea that each base in a read can with high probability be predicted by the context of bases that are next to it.
We also exploit the fact that such predicted bases add little information, and its quality score can be discarded or heavily compressed without distortion effects on downstream analysis. 

In our strategy, the length of contexts is {\it variable-order} (\ie not fixed {\it a priori}, unlike  BEETL), and can be as large as the full read length for high-enough coverages and small-enough error rates.

We exploit the positional clustering framework introduced in  \cite{prezza2019snps} to detect ``relevant'' blocks in the \EBWT. 
These blocks allow us not only to smooth the quality scores, but also to apply a noise reduction on the corresponding bases, replacing those that are believed to be noise, 
while keeping variant calling performance comparable to that with the original data and with compression rates comparable to other tools. 

\begin{table}[t]
\caption{{Extended Burrows-Wheeler Transform (EBWT), LCP array, and auxiliary} data structures used for detecting positional clusters for the set $\bigS=\{GGCGTACCA\$_1, GGGGCGTAT\$_2, ACGANTACGAC\$_3\}$ and $k_m=2$. }
\begin{center}
\scriptsize
\centering
\begin{tabular}{@{ }c@{ }c@{ }c@{ }c@{ }c@{ }|c@{}c@{}c@{}c@{}c@{}c@{}c@{}c@{}c@{}c@{}c@{}c@{}cc@{ }c@{ }c@{ }c@{ }c@{ }|c@{}c@{}c@{}c@{}c@{}c@{}c@{}c@{}c@{}c@{}}
$i$&$\LCPVmin$&$\LCPVthr$&$\lcp$&$\ebwt$ & \multicolumn{12}{c}{\mbox{Sorted Suffixes}} &         & $i$  &$\LCPVmin$ &$\LCPVthr$&$\lcp$&$\ebwt$ & \multicolumn{10}{c}{\mbox{Sorted Suffixes}} \\
\cline{1-17} 
\cline{19-31} 
1 &0 & {0} & {0} & {A} & {$\$_1$} &  &  &  &  &  &  &  &  &  &  &  &                & 17 &0 & 1 & 4 & {G} & C & G & T & A & T & $\$_2$ &  &  &  &  \\
2 &0&   0 & 0   & T   & $\$_2$   &  &  &  &  &  &  &  &  &  &  &  &                & 18 &1& 0 & 0 & C & G & A & C & $\$_3$ &  &  &  &  &  &  \\
3 &0& 0 & 0 & C & $\$_3$         &  &  &  &  &  &  &  &  &  &  &  &                & 19 &0& 1 & 2 & C & G & A & N & T & A & C & G & A & C & $\$_3$ \\
4 &0& {0} & {0} & {C} & {A} & {$\$_1$} &  &  &  &  &  &  &  &  &  &  &             & 20 &1& {0} & {1} & {G} & {G} & {C} & {G} & {T} & {A} & {C} & {C} & {A} & {$\$_1$} &  \\
5 &0 & 0 & 1 & {G} & A & C & $\$_3$ &  &  &  &  &  &  &  &  &  &                    & 21 &0 & 1 & 5 & G & G & C & G & T & A & T & $\$_2$ &  &  &  \\
6 &0& {1} & {2} & {{T}} & {A} & {C} & {C} & {A} & {$\$_1$} &  &  &  &  &  &  &  &  & 22 &1 & {0} & {1} & {{\$}} & {G} & {G} & {C} & {G} & {T} & {A} & {C} & {C} & {A} & {$\$_1$} \\
7 &0& 1 & 2 & {T} & A & C & G & A & C & $\$_3$ &  &  &  &  &  &  &                 & 23 &0 & 1 & 6 & {G} & G & G & C & G & T & A & T & $\$_2$ &  &  \\
8 &0& 1 & 4 & {\$} & A & C & G & A & N & T & A & C & G & A & C & $\$_3$ &          & 24 &1 & 1 & 2 & {G} & G & G & G & C & G & T & A & T & $\$_2$ &  \\
9 &1& 0 & 1 & G & A & N & T & A & C & G & A & C & $\$_3$ &  &  &  &                & 25 &0 & 1 & 3 & {\$} & G & G & G & G & C & G & T & A & T & $\$_2$ \\
10 &0& 0 & 1 & T & A & T & $\$_2$ &  &  &  &  &  &  &  &  &  &                     & 26 &1 & {0} & {1} & {C} & {G} & {T} & {A} & {C} & {C} & {A} & {$\$_1$} &  &  &  \\
11 &1& 0 & 0 & A & C & $\$_3$ &  &  &  &  &  &  &  &  &  &  &                      & 27 &0 & 1 & 3 & C & G & T & A & T & $\$_2$ &  &  &  &  &  \\
12 &0& {0} & {1} & {C} & {C} & {A} & {$\$_1$} & {} &  &  &  &  &  &  &  &  &       & 28 &1 & 0 & 0 & A & N & T & A & C & G & A & C & $\$_3$ &  &  \\
13 &0& {0} & {1} & {A} & {C} & {C} & {A} & {$\$_1$} &  &  &  &  &  &  &  &  &      & 29 &0 & 0 & 0 & A & T & $\$_2$ &  &  &  &  &  &  &  &  \\
14 &0& 0 & 1 & {A} & C & G & A & C & $\$_3$ &  &  &  &  &  &  &  &                 & 30 &0 & {0} & {1} & {{G}} & {T} & {A} & {C} & {C} & {A} & {$\$_1$} &  &  &  &  \\
15 &0& 1 & 3 & {A} & C & G & A & N & T & A & C & G & A & C & $\$_3$ &  &           & 31 &0 & 1 & 3 & {N} & T & A & C & G & A & C & $\$_3$ &  &  &  \\
16 &1 & 1 & 2 & G & C & G & T & A & C & C & A & $\$_1$ &  &  &  &  &                & 32 &0 & 1 & 2 & {G} & T & A & T & $\$_2$ &  &  &  &  &  &
\end{tabular}
\end{center}
\label{table:eBWTexe3seqs}
\end{table}

\section{Preliminaries}

Let $S$ be a string (also called sequence or {\it reads} due to our target application) of length $n$ on the alphabet $\Sigma$. 
We denote the $i$-th symbol of $S$ by $S[i]$. 
A {\it substring} of any $S \in \bigS$ is denoted as $S[i,j] = S[i] \cdots S[j]$, with $S[1,j]$ being called a {\it prefix} and $S[i,n+1]$ a {\it suffix} of $S$.

Let $\bigS=\{S_1,S_2,\ldots,S_{m}\}$ be a collection of $m$ strings.
We assume that each string $S_i \in \bigS$ has length $n_i$ and is followed by a special end-marker symbol $S_i[n_i+1] = \$_i$, which is lexicographically smaller than any other symbol in $\bigS$, and does not appear in $\bigS$ elsewhere.

The Burrows-Wheeler Transform (\BWT) \cite{bwt94} of a text $T$ (and the \EBWT of a set of strings $\bigS$ \cite{MantaciRRS07,BauerCoxRosoneTCS2013})
is a suitable permutation of the symbols of $T$ (and $\bigS$) whose output shows a local similarity, 
\ie symbols preceding similar contexts tend to occur in clusters.
Both transformations have been intensively studied from a theoretical and combinatorial viewpoint and 
have important and successful applications in several areas, \eg ~\cite{MantaciRRS08,LiDurbin10,KKbioinf15,YALFF2019,GagieNP20,GuerriniLouzaRosone2020}.

We assume that $N=\sum_{i=1}^{m}(n_i+1)$ denotes the sum of the lengths of all strings in $\bigS$.
The output of the \EBWT is a string $\ebwt(\bigS)$ of length $N$ such that 
$\ebwt(\bigS)[i]=x$, with $1\!\leq\!i\!\leq\!N$, if $x$ circularly precedes the $i$-th suffix (context) $S_j[k,n_j+1]$ (for some $1\leq j\leq m$ and $1\leq k\leq n_j\!+\!1$), according to the lexicographic sorting of the contexts of all strings in $\bigS$. 
In this case, we say that the context $S_j[k,n_j+1]$ is associated with the position $i$ in $\ebwt(\bigS)$.
See Table~\ref{table:eBWTexe3seqs} for an example.

In practice, computing the \EBWT via suffix sorting \cite{BauerCoxRosoneTCS2013,BoMaReRoSc_IJFCS_2014}
may be done considering the same end-markers for all strings.
That is, we assume that $\$_i < \$_j$, if $i<j$, and use a unique symbol $\$$ as the end-marker for all strings.

The {\it longest common prefix} ($\LCP$) array~\cite{Manber:1990}
of  $\bigS$ is the array $\lcp(\bigS)$ of length $N+1$, such that for $2 \leq i \leq N$, $\lcp(\bigS)[i]$ is the length of the longest common prefix between the 
contexts associated with the positions $i$ and $i-1$ in $\ebwt(\bigS)$, while $\lcp(\bigS)[1] = \lcp(\bigS)[N+1]= 0$.
The $\LCP$-intervals \cite{KurtzOhlebusch2004} are maximal intervals $[i,j]$ that satisfy $\lcp(\bigS)[r] \geq k$ for {$i < r \leq j$} and whose associated contexts share at least the first $k$ symbols.

An important property of the \BWT and \EBWT, 
is the so-called {\it LF mapping}~\cite{FerraginaManzini2000}, which states that the $i$-th occurrence of symbol $x$ on the BWT string and the first symbol of the $i$-th lexicographically-smallest suffix that starts with $x$ correspond to the same position in the input string (or string collection).
We will use the LF mapping to perform
{\it backward searches} when creating a new (modified) FASTQ file.
The backward search allows to find the range of suffixes prefixed by a given string -- see \cite{FerraginaManzini2000,bookBWTAdjeroh:2008} for more details.

\section{Method}

We structure our reference-free FASTQ compression method in four
main steps: 
(a) data structures building, 
(b) positional cluster detecting, (c) noise reduction and quality score smoothing, and (d) FASTQ reconstruction. 
\paragraph{\bf(a) Data structures building.}
This phase consists in computing the \EBWT and the \LCP array for the collection of sequences $\bigS$ stored in the {\it bases} component of the input FASTQ file.
We also compute $\qs(\bigS)$, as the concatenation of the quality scores associated with each symbol in $\ebwt(\bigS)$, \ie, the string $\qs$ contains a permutation of the quality score symbols that follows the symbol permutation in $\ebwt(\bigS)$. 
Note that the $\lcp(\bigS)$ is only used in the next step, so we can either explicitly compute it in this phase or implicitly deduce it by $\ebwt(\bigS)$ during the next step.

\paragraph{\bf (b) Positional cluster detecting.}
A crucial property of the \EBWT is 
that symbols preceding suffixes that begin with the same substring (context) $w$ will result in a contiguous substring of $\ebwt$, and thus of $\qs$. Such a substring of the $\ebwt$ is generally called {\it cluster}.
In literature, such clusters depending on the length $k$ of the context $w$ are associated with $\LCP$-intervals \cite{KurtzOhlebusch2004}.

The aim of the positional clustering framework~\cite{prezza2019snps,PrezzaPisantiSciortinoRosone2020} is to overcome the limitation of strategies based on \LCP-intervals, which depend on the choice of $k$. Intuitively, 
meaningful clusters in the \EBWT lie between local minima in the \LCP array, 
and symbols of the same positional cluster usually cover the same genome location~\cite{prezza2019snps}.
This recent strategy automatically detects, in a data-driven way, the length $k$ of the common prefix shared by the suffixes of a cluster in the \EBWT. 
Moreover, so as to exclude
clusters corresponding to short random contexts, we set a minimum length for the context  $w$, denoted by $k_m$. 

Analogously to~\cite{PrezzaPisantiSciortinoRosone2020}, we define positional clusters by using two binary vectors: $\LCPVthr$ and $\LCPVmin$, where $\LCPVthr[i]=1$ if and only if $\lcp[i]\geq k_{m}$, 
and $\LCPVmin=1$ if and only if $\lcp[i]$ is a local minimum {\it i.e.}, it holds $\lcp[i-1]>\lcp[i]\leq\lcp[i+1]$, for all $1<i\leq N$, which depends on data only.
A \EBWT positional cluster is a maximal substring $\ebwt[i,j]$ such that $\LCPVthr[r] = 1$, for all $i< r \leq j$, and $\LCPVmin[r] = 0$, for all $i < r \leq j$.
See, for instance, Table~\ref{table:eBWTexe3seqs}.

\paragraph{\bf (c.1) Noise reduction.}
Given the base symbols appearing in any positional cluster of $\ebwt(\bigS)$, we call as {\it frequent symbol} any symbol whose occurrence in the cluster is greater than a threshold percentage. 
The idea that lies behind changing bases is to reduce the number of symbols in a cluster that are different from the most frequent symbols, while preserving the variant calls.
So, we take into account only clusters that have no more than two frequent symbols (for example, we set the threshold percentage to $40$\%).

The symbols in an \EBWT positional cluster usually correspond to the same genome location \cite{prezza2019snps}. 
Thus, given an \EBWT positional cluster $\alpha~=\ebwt[i,j]$, we say a symbol $b$ is a {\it noisy base} if
it is different from the most frequent symbols and all occurrences of $b$ in $\alpha$ are associated with low quality values in {$\qs[i,j]$} (\ie there are no occurrences of $b$ with a high quality score in $\alpha$). 
Intuitively, a noisy base is more likely noise introduced during the sequencing process.

Then, for each analyzed cluster $\alpha$, we replace noisy bases in $\alpha$ with a predicted base $c$ as follows.
We distinguish two cases.
If the cluster $\alpha$ contains a unique most frequent symbol $c$, then we replace the noisy base $b$ with $c$.
Otherwise, if we have two different frequent symbols, for each occurrence of them and for the noisy base $b$,
we compute the preceding context of length $\ell$ 
in their corresponding reads (\ie left context of each considered base), 
by means of the backward search applied to $\ebwt(\bigS)$
(for example, we set $\ell=1$ in our experiments).
If the left context preceding $b$
coincides with the contexts preceding all the occurrences of $d$
(one of the two most frequent symbols), then we replace the base $b$ with $d$. 
We specify that no base changes are performed if 
the frequent symbols are preceded by the same contexts.

\paragraph{\bf (c.2) Quality score smoothing.}
During step (c.1), we also modify quality scores
by smoothing the symbols of $\qs$ that 
are associated with base symbols in clusters of $\ebwt$.

In any cluster $\alpha$, the value $Q$ used for replacements can be computed with different strategies:
(i) $Q$ is a default value, or 
(ii) $Q$ is the quality score associated with the mean probability error in $\alpha$, or 
(iii) $Q$ is the maximum quality score in $\alpha$, or 
(iv) $Q$ is the average of the quality scores in $\alpha$.
According to this smoothing process, apart from strategy (i), 
the value $Q$ depends on the cluster analyzed.
In the experiments we evaluated $qs$ smoothing as follows.
For each position $r$ within $\alpha=\ebwt[i,j]$
we smooth the quality score $qs[r]$ with $Q$, either if $\ebwt[r]$ is one of the most frequent symbols (regardless of its quality), or if $qs[r]$ is greater than $Q$.

An additional feature to compress further quality scores is the possibility of reducing the number of the alphabet symbols appearing in $\qs(\bigS)$. This smoothing approach is quite popular and standard in literature~\cite{SPRING2018}.
Then, in addition to one of the 
strategies described above, we can apply the Illumina 8-level binning 
reducing to $8$ the number of different symbols in $\qs$.

\paragraph{\bf (d) FASTQ reconstruction and compression.}
{
Given the modified symbols in
$\ebwt(\bigS)$ and in $\qs(\bigS)$  (according to the strategies described above), we use the LF mapping on the original $\ebwt$ string to retrieve the order of symbols and output a new (modified) FASTQ file.
}

{
The {\it headers} component with the read titles can be either omitted (inserting the symbol `@' as header) or kept as they are in the original FASTQ file.
}

At the end, the resulting FASTQ file is compressed by using any state-of-the-art compressor.
    
\begin{table}[t]
\begin{center}
\caption{Paired-end datasets used in the experiments and their sizes in bytes. Each dataset is obtained from two files (\_1 and \_2), whose number of reads and read length are given in columns 2 and 3. 
We distinguish the size of the original FASTQ (raw data) from the size of the same FASTQ file with all headers removed ({\it i.e.}, replaced by '@'). In the last column we report the size of the bases component, 
that is equal to the quality scores component size.}
\scriptsize{
\begin{tabular}{|c|c|c|c|c|c|}
\hline  
Dataset & No. reads & Length   & Raw (complete)    & FASTQ  & DNA/QS  \\
\hline  
ERR262997\_1 - chr 20\, &\, 13,796,697\, & 101  & 3,420,752,544     & \, 2,869,712,976\,    & \, 1,407,263,094\, \\
ERR262997\_2 - chr 20\, & \, 13,796,697\, & 101 & 3,420,752,544     & \, 2,869,712,976\,    & \, 1,407,263,094\, \\
\hline  
ERR262997\_1 - chr 14\, & \, 18,596,541\,& 101  & 4,611,888,574     & \, 3,868,080,528 \,   & \, 1,896,847,182\, \\
ERR262997\_2 - chr 14\, & \, 18,596,541\,& 101  & 4,611,888,574     & \, 3,868,080,528 \,   & \, 1,896,847,182\, \\
\hline  
ERR262997\_1 - chr 1\phantom{0}\, & \, 49,658,795\,& 101   & 12,211,743,094    & \,10,329,029,360 \,   &\, 5,065,197,090\, \\
ERR262997\_2 - chr 1\phantom{0}\, & \, 49,658,795\,& 101   & 12,211,743,094    & \,10,329,029,360 \,   &\, 5,065,197,090\, \\
\hline
\end{tabular}
\label{tabDataset}
}
\end{center}
\end{table}

\section{Experiments}\label{sec:experiments}

\paragraph{\bf Implementations.}
We present two implementations of our tool \fastQC:
in internal memory and in external memory. Now we give a brief description of them.

Given a FASTQ file containing a collection $\bigS$, both implementations take as input the files containing $\ebwt(\bigS)$ and $\qs(\bigS)$. In the external memory, we also need the array $\lcp(\bigS)$.

The construction of these data structures during the first step can be performed with any tools. 
\eg~\cite{BauerCoxRosoneTCS2013,Bonizzoni2019,EgidiLMT19,LouzaTGPrezzaRosone2020,SciortinoSPIRE2021} according to the resources available (which is a good feature).

The two implementations largely differ in the detection of positional clusters. Indeed, alike~\cite{PrezzaPisantiSciortinoRosone2020}, the internal memory approach represents $\ebwt(\bigS)$ via the compressed suffix tree described in~\cite{PR_TCS2021} (see also \cite{Belazzougui2020}), where it is shown that $\lcp(\bigS)$ can be induced from the \EBWT using succinct working space for any alphabet size. 
Whereas, in external memory, \EBWT positional clusters are detected by reading the $\lcp(\bigS)$ stored in a file in a sequential way. 

For the last two steps, the two implementations are similar, except that the data are kept in internal or external memory.
In particular, during step (d), we use the LF-mapping either in internal memory -- via the suffix-tree navigation as in \cite{PrezzaPisantiSciortinoRosone2020} --
or in external memory -- similarly to \cite{BauerCoxRosoneTCS2013}.

\paragraph{\bf Datasets.}
The easiest approach to evaluate the validity of our method could be to simulate reads and variants from a reference genome. However, it is not trivial to simulate variant artifacts for this purpose \cite{Li2013Toward}, so we focus only on real data.

In this study, thus, we use the real human dataset ERR262997 corresponding to a 30x-coverage paired-end Whole Genome Sequencing (WGS) data for the CEPH 1463 family. Similarly to other studies \cite{OchoaHernaezGoldfederWeissmanAshley2016}, for evaluation purposes we extracted the chromosomes $20$, $14$ and $1$ from ERR262997, obtaining datasets of different sizes (see Table \ref{tabDataset}).

\paragraph{\bf Compression.}

We describe experiments that show that our strategy is able to compress FASTQ files in lossy way modifying both the bases and quality scores component, while keeping most of the information contained in the original file.

\begin{table}[t]
{\scriptsize
\begin{center}
\caption{\label{tabCompression}Compression ratio for (original and three smoothing tools)  FASTQ files (with headers replaced by '@') and their single components (qualities (QS) and bases (DNA))
obtained by both PPMd and BSC. The ratio is defined as compressed size/original size, where original file size is in Table \ref{tabDataset}. 
Since BEETL and LEON do not modify the bases component, their ratio for the DNA component is the same as the original. 
}
\begin{tabular}{|c@{\ }|c|c|@{ \ \ \ }c@{ \ \ \ }|@{ \ \ \ }c@{ \ \ \ }||c|@{ \ \ \ }c@{ \ \ \ }|@{ \ \ \ }c@{ \ \ \ }|}
\hline
\multirow{2}{*}{}     &  \multirow{2}{*}{Tool}     & ~FASTQ~  & QS  & DNA  & ~FASTQ~     & QS      & DNA     \\
\cline{3-8}\addlinespace[0.06em]
                   &   & \multicolumn{3}{c||}{ERR262997\_1  chr 20} & \multicolumn{3}{c|}{ERR262997\_2 chr 20} \\
\hline
\multirow{4}{*}{PPMd} & Original  & 0.2473  & 0.2999   & \multirow{3}{*}{0.2037}    & 0.2547     & 0.3142       & \multirow{3}{*}{0.2046} \\
                      & LEON      & 0.1152  & 0.0317    &    & 0.1234     & 0.0473       &         \\
                      & BEETL     & 0.1900  & 0.1833   &     & 0.2002     & 0.2033       &         \\
\cline{2-8}\addlinespace[0.06em]          
                      & \fastQC  & 0.1941  & 0.1918   & 0.2035    & 0.2043    & 0.2119       & 0.2043        \\
\hline
\multirow{4}{*}{BSC}  & Original  & 0.2005  & 0.2902   & \multirow{3}{*}{0.1154}    & 0.2095     & 0.3034       & \multirow{3}{*}{0.1207} \\
                      & LEON      & 0.0677  & 0.0241   &    & 0.0780     & 0.0394       &        \\
                      & BEETL     & 0.1413  & 0.1724   &    & 0.1534     & 0.1915       &        \\
\cline{2-8}\addlinespace[0.06em]
                      & \fastQC  & 0.1453  & 0.1830   & 0.1152    & 0.1572     & 0.2024       & 0.1200        \\
\hline
\hline
    \multicolumn{2}{|c}{}                 & \multicolumn{3}{|c||}{ERR262997\_1  chr 14} & \multicolumn{3}{c|}{ERR262997\_2 chr 14} \\
\hline
\multirow{4}{*}{PPMd} & Original  & 0.2482  & 0.2956   & \multirow{3}{*}{0.2100}    & 0.2544     & 0.3076       & \multirow{3}{*}{0.2106} \\
                      & LEON      & 0.1175  & 0.0301   &     & 0.1249     & 0.0444       &         \\
                      & BEETL     & 0.1916  & 0.1805   &     & 0.2010     & 0.1989       &        \\
\cline{2-8}\addlinespace[0.06em]          
                      & \fastQC  & 0.1957  & 0.1889   & 0.2098    & 0.2050     & 0.2074       & 0.2103        \\
\hline
\multirow{4}{*}{BSC}  & Original  & 0.1992  & 0.2862   & \multirow{3}{*}{0.1174}    & 0.2071     & 0.2972       & \multirow{3}{*}{0.1224} \\
                      & LEON      & 0.0674  & 0.0226   &     & 0.0770     & 0.0367       &        \\
                      & BEETL     & 0.1406  & 0.1698   &     & 0.1518     & 0.1874       &  \\
\cline{2-8}\addlinespace[0.06em]
                      & \fastQC  & 0.1445   & 0.1786   & 0.1164    & 0.1555     & 0.1962       & 0.1210  \\
\hline
\hline
\multicolumn{2}{|c|}{}                 & \multicolumn{3}{c||}{ERR262997\_1  chr 1} & \multicolumn{3}{c|}{ERR262997\_2 chr 1} \\
\hline
\multirow{4}{*}{PPMd} & Original   & 0.2461   & 0.2969  & \multirow{3}{*}{0.2046}  & 0.2529       & 0.3097      & \multirow{3}{*}{0.2054}      \\
                      & LEON       & 0.1148   & 0.0299  &   & 0.1224       & 0.0445 &  \\
                      & BEETL      & 0.1876   & 0.1777  &  & 0.1973       & 0.1966  &  \\
\cline{2-8}\addlinespace[0.06em]
                      & \fastQC   & 0.1918   & 0.1864  & 0.2044 & 0.2015       & 0.2054      & 0.2051      \\
\hline
\multirow{4}{*}{BSC}  & Original   & 0.1984   & 0.2874  & \multirow{3}{*}{0.1146}  & 0.2068       & 0.2991      & \multirow{3}{*}{0.1197}      \\
                      & LEON       & 0.0661   & 0.0224  &  & 0.0758       & 0.0367      &  \\
                      & BEETL      & 0.1379   & 0.1669  &  & 0.1494       & 0.1850  & \\
\cline{2-8}\addlinespace[0.06em]
                      & \fastQC   & 0.1419   & 0.1759  & 0.1136  & 0.1533       & 0.1941      & 0.1183     \\
\hline
\end{tabular}
\end{center}
}
\end{table}

To the best of our knowledge, none of the existing FASTQ compressors evaluates and modifies both the {\it bases} and the {\it quality scores} components at the same time.
Therefore,
no comparison with existing tools is
completely fair.
We consider the tools BEETL\footnote{\url{https://github.com/BEETL/BEETL/blob/RELEASE_1_1_0/scripts/lcp/applyLcpCutoff.pl}} \cite{JaninRosoneCox2014}
and LEON\footnote{\url{http://gatb.inria.fr/software/leon/}} \cite{LEON2015}
that are reference-free and read-based:
they smooth the {\it quality score} component in a lossy way based on the biological information of the {\it bases}, 
without modifying  the {\it bases} themselves.

We choose BEETL, because it is based on \EBWT and takes as input the same data structures we compute during step (a). 
BEETL smooths to a constant value the quality scores corresponding to each run of the same symbol associated with the $\LCP$-interval~$[i,j]$.
The quality scores associated to each run in $\ebwt[i,j]$ are smoothed if the length of the run is greater than a minimum stretch length $s$. So, it needs two parameters: the minimum stretch length $s$ and the cut threshold $c$ for the LCP-interval, failing to separate read suffixes differing after $c$ positions (this is, indeed, a drawback of all $c$-mer-based strategies). 

We choose LEON \cite{LEON2015} that, on the contrary, needs to build a reference from the input reads in the form of a bloom filter compressed {\it de Bruijn} graph, and then maps each nucleotide sequence as a path in the graph. 
Thus, LEON can be considered assembly-based, since it uses a {\it de Bruijn} graph as a de novo reference.
If a base is covered by a sufficiently large number of $k$-mers (substrings of length $k$) stored in the bloom filter, its quality is set to a fixed high value. 
Thus, LEON depends on a fixed parameter $k$ for the graph, as well.

\begin{table}[t]
\begin{center}
\caption{\label{table:vsOriginal_BWA_GATK}Evaluation of called variants by means of {\tt rtg vcfeval}: comparison between called variants from a modified FASTQ and 
variants from the original FASTQ used as baseline.}
\scriptsize{%
\begin{tabular}{|c|c|c|c||c|c|c||c|c|c|}
\hline
         & \multicolumn{3}{c||}{ERR262997 chr 20}                         & \multicolumn{3}{c||}{ERR262997 chr 14}                         & \multicolumn{3}{c|}{ERR262997 chr 1}     \\ 
         \hline
         & \PREC       & \SEN     & \F1       & \PREC       & \SEN     & \F1       & \PREC       & \SEN     & \F1       \\ \hline
LEON     & 0.9593          & 0.9356          & 0.9473          & 0.9626          & 0.9381          & 0.9502          & 0.9589          & 0.9348          & 0.9467          \\ \hline
BEETL    & 0.9584          & 0.9529          & 0.9556          & 0.9626          & 0.9553 & 0.9590          & 0.9596          & \textbf{0.9526} & 0.9561          \\ \hline
~\fastQC~ & \textbf{0.9613} & \textbf{0.9534} & \textbf{0.9574} & \textbf{0.9650} & \textbf{0.9555} & \textbf{0.9602} & \textbf{0.9628} & 0.9523          & \textbf{0.9575} \\ \hline
\end{tabular}%
}
\end{center}
\end{table}

Regarding the output of each tool, \fastQC
produces a new FASTQ file with modified bases and smoothed qualities, whereas BEETL only produces a BWT-ordered smoothed qualities file, 
which is used to replace the quality scores component in the modified FASTQ file (with the original bases component).
LEON produces a proprietary format compressed file that encodes the {\it de Bruijn} graph, thus, it was necessary to uncompress the output file 
to obtain the modified FASTQ.
For a fairer comparison,
these resulting FASTQ files were compressed with the same tools.
In particular, we choose two well-known compressors for this task: PPMd
\cite{PPM1984,PPM1990} and BSC\footnote{\url{http://libbsc.com/}}.

We run \fastQC
and BEETL with similar parameters: in BEETL, we set the replacement quality score to `$@$' (as set by LEON) and the minimum \LCP cut threshold to $30$.
For \fastQC,  we used options {\tt-T 30} to set the minimum context length $k_m=30$ and {\tt -Q @} to set the constant replacement value. 
LEON was executed with default parameters for $k$-mer size and minimal abundance threshold, as suggested by the authors.
The exact commands for the tools are:
\begin{itemize}
    \item \texttt{python3 BFQzip.py { <input>}.fastq -o {<output>}.fastq -T 30 -Q @}
    \item \texttt{applyLcpCutoff.pl -b {<input>}.ebwt -q {<input>}.ebwt.qs -l {<input>}.lcp -o {<output>}.ebwt.qs -c 30 -r 64 -s 5}
    \item \texttt{leon -file {<input>}.fastq -c ({\rm for decompression} -d) -nb-cores 1}
\end{itemize}

In Table \ref{tabCompression}, we report the compression ratios achieved by PPMd and BSC given as input the FASTQ modified by any of the three tools and the original FASTQ file. Note that each of the two FASTQ files comprising any paired-end dataset is compressed separately.

Table \ref{tabCompression} shows that all tools improve the compression of the data (compared with the original FASTQ).
In particular, LEON is better in terms of compression ratios than other tools. 
This improvement is due to a greater ability to smooth the quality scores component. Recall that LEON truncates all quality scores above a given threshold (qualities higher than ‘@’ are replaced by ‘@’) and in these datasets the total frequency of symbols greater than ‘@’ is about $80-90$\%.
However, the results of \fastQC and BEETL were similar in almost all cases. 

\paragraph{\bf Validation.}

\begin{figure}[t!]
         \centering
        \includegraphics[width=1\textwidth]{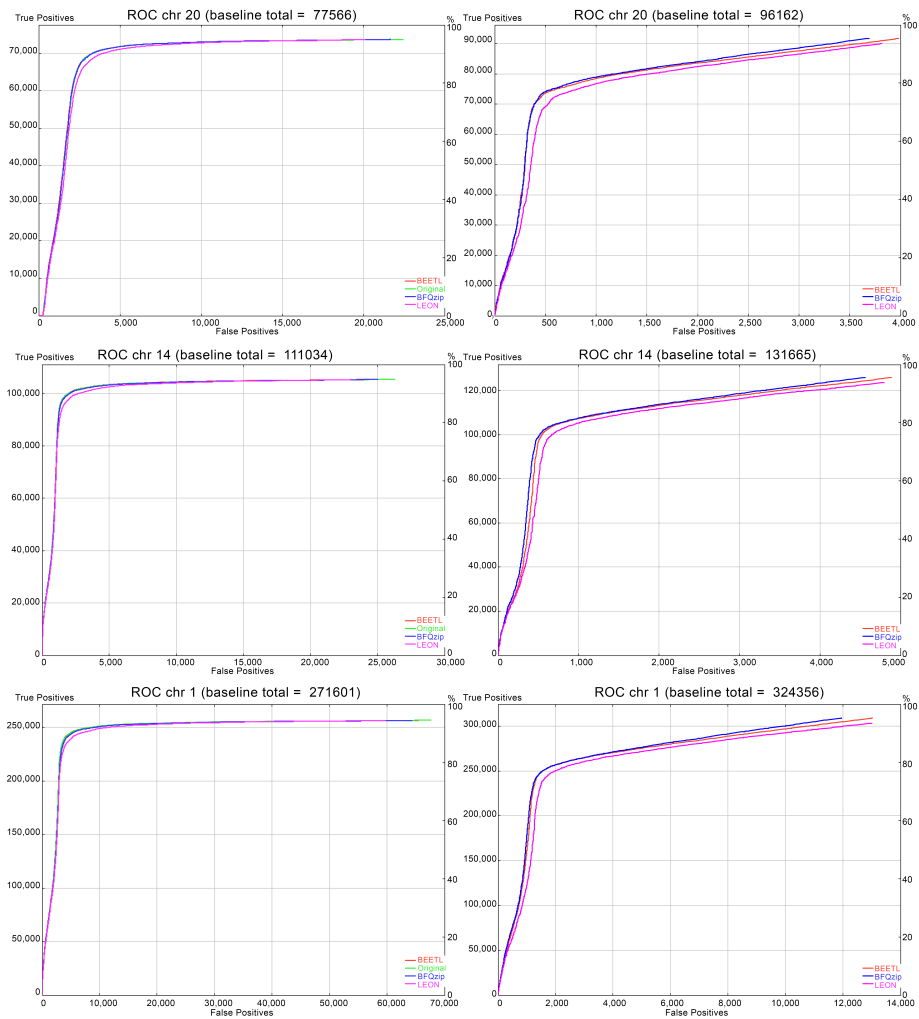}
         \caption{ROC Curves obtained by \texttt{rtg rocplot}:
         true positive as a function of the false positive respect both to the 
         Ground truth as baseline (left side) and to the
         original file as baseline (right side)}
         \label{fig:allChr_TPFP}
\end{figure}

\begin{figure}[t!]
         \centering
        \includegraphics[width=1\textwidth]{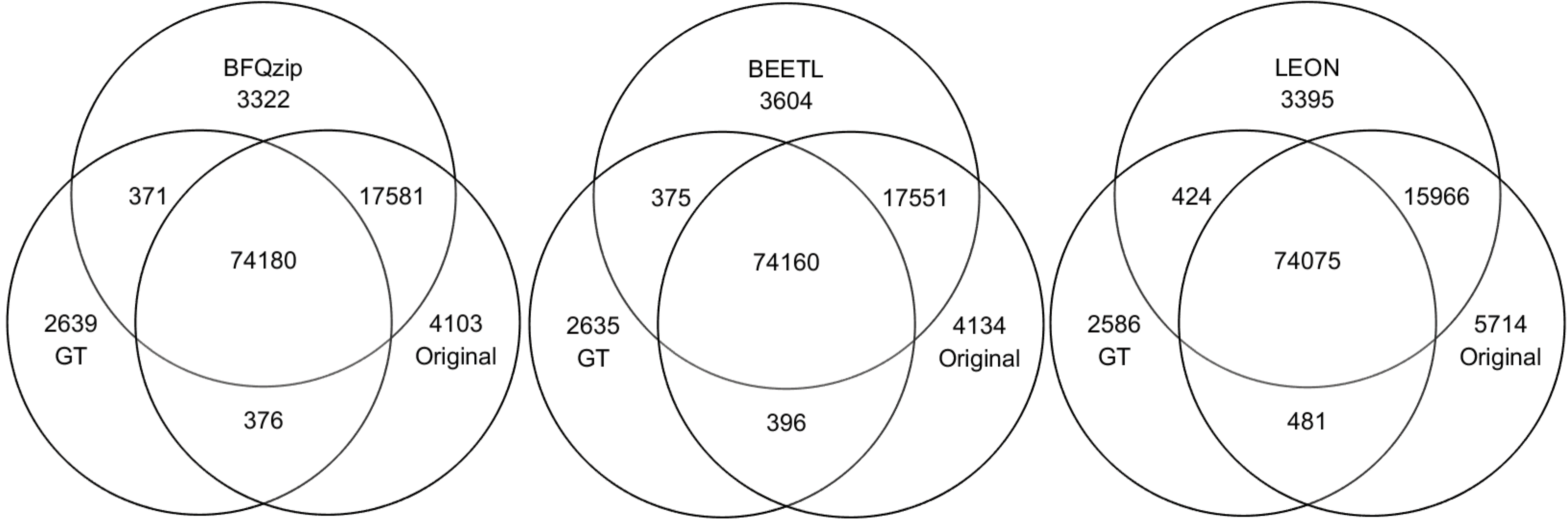}
       \caption{Venn diagrams for chr 20 variants: set comparison between the variants in the original FASTQ file (Original), those in the ground truth (GT) and those called from the FASTQ modified by BFQzip (left), or modified by BEETL (middle), or modified by LEON (right).}
        \label{fig:chr20VennDiagram}
\end{figure}

\begin{figure}[t!]
         \centering
        \includegraphics[width=1\textwidth]{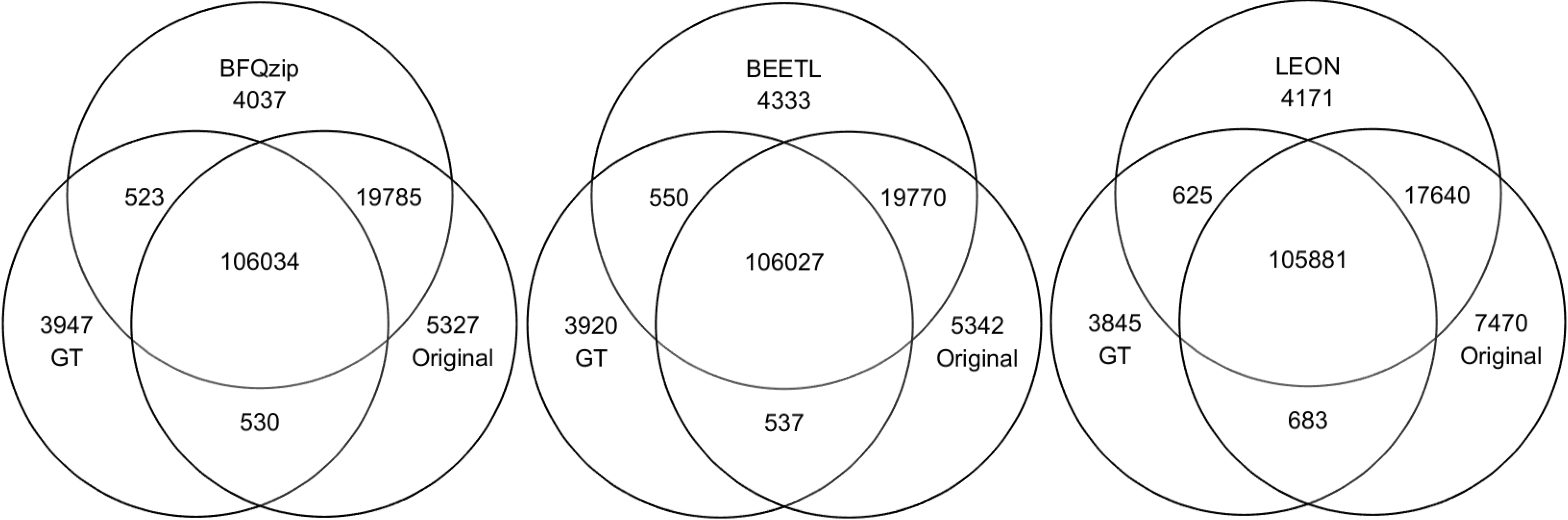}
       \caption{Venn diagrams for chr 14 variants: set comparison between the variants in the original FASTQ file (Original), those in the ground truth (GT) and those called from the FASTQ modified by BFQzip (left), or modified by BEETL (middle), or modified by LEON (right).}
        \label{fig:chr14VennDiagram}
\end{figure}


In lossy FASTQ compression, 
it is important to take
into account the impact of the modified data on downstream analysis, so we need to evaluate the genotyping accuracy.

We compared the set of variants retrieved from a baseline with the set retrieved from the modified FASTQ.
First, we considered as baseline the set of ``ground truth'' variants for NA12878 provided by Illumina\footnote{\url{https://github.com/Illumina/PlatinumGenomes}} and then, we considered as baseline the set of variants obtained from the original FASTQ files.

The SNP calling pipeline is a bash script\footnote{\url{https://github.com/veronicaguerrini/BFQzip/blob/main/variant_calling/pipeline_SNPsCall.sh}} \cite{li2013aligning} 
to align sequences to the reference (in our case, the latest build of the human reference genome, GRCh38/hg38) and GATK-HaplotypeCaller \cite{DePristoGATK} to call SNPs. The output of the pipeline is a VCF file that contains the set of called SNPs, which are then compared
to those contained in a baseline set
by using RTG Tools\footnote{\url{https://www.realtimegenomics.com/products/rtg-tools}}: \texttt{rtg vcfeval} evaluates agreement between called and baseline variants using the following performance metrics.

True positive ($TP$) are those variants that match in both  baseline and query (the set of called variants); false positives ($FP$) are those variants that have mismatching, that are in the called set of variants but not in the baseline;
false negatives ($FN$) are those variants that are present in the baseline, but missing in the query. 
These values are employed to compute precision ($\PREC$) that measures the proportion of called variants that are true, and sensitivity ($\SEN$), which measures the proportion of called variants included in the consensus set. A third metric 
is the harmonic mean between sensitivity and precision (known as $\F1$-measure):
\[\PREC=\frac{TP}{TP+FP}\,,\qquad\SEN=\frac{TP}{TP+FN}\,,\]
\[\F1=\frac{2\cdot SEN\cdot PREC}{SEN+PREC}\,.\]

The tool \texttt{rtg vcfeval} can also output a ROC file based on the \texttt{QUAL} field, which can be viewed by \texttt{rtg rocplot}.


Table \ref{table:vsOriginal_BWA_GATK} reports the evaluation results of the variants retrieved from a modified FASTQ (produced by any tool) compared to the set of variants from the original (unsmoothed) FASTQ, which is used as baseline. 
We observe that our method provides the highest precision and F measure.
This means that, with respect to the original FASTQ, we have a higher number of common variants (TP) and the lowest number of newly-introduced variants (FP), thus we preserve information of the original FASTQ more than the others (see also Figure \ref{fig:allChr_TPFP}, right side).
This is a desirable property useful in several applications.

In Figure \ref{fig:allChr_TPFP} we provide the ROC curves associated with the variant comparison performed by {\tt rtg vcfeval} (reporting the number of TP as a function of the FP), showing similar results.

Figure \ref{fig:allChr_TPFP} (left side) shows that our tool preserves those variants that are in the ground truth: using the VCF file of the ground truth as baseline, all tools show an accuracy similar to the original. In particular, \fastQC and BEETL curves in each graph overlap the original one, while LEON's curve is a bit lower.

Figure \ref{fig:allChr_TPFP} (right side) also
shows that our tool preserves the original variants (TP) introducing only a little number of new ones (FP) when the original FASTQ variants are used as baseline (as also shown by the percentages of sensitivity and precision in Table \ref{table:vsOriginal_BWA_GATK}) .

With the idea of showing the preservation of variants from the original file and the possibility of losing variants due to the sequencer noise, we decided to manually check the variants obtained for any tool by intersecting the corresponding vcf file with that from the FASTQ original and with the ground truth vcf.
Figures \ref{fig:chr20VennDiagram} and \ref{fig:chr14VennDiagram} show as our tool preserves the variants that are both in the original file and in the ground truth (\ie GT) more than the others.

This analysis appears to confirm what we observed with the ROC curves.
\fastQC reports the smaller number of new variants introduced by the tool, that are those variants neither in the original file nor in the ground truth.

The majority of the variants present in both the original FASTQ and the ground truth is preserved by all tools. However, it appears that the heavy truncation of the quality scores carried out by LEON leads to a loss of variants present in the original file (see, for instance, the intersection between each tool and the original, or the intersection between the ground truth and the original). 
While \fastQC (and in the similar way BEETL) preserves a high number of variants that are present in the original file (see intersection between each tool and the original).

We also made a detailed analysis of the modifications performed by our method, comparing the modified bases which correspond to parts aligned by BWA with the relative bases in the reference (taking into account how BWA aligned the reads). We have noticed that about $91$-$93$\% of the changed bases correspond to the reference. The remaining part includes positions we cannot evaluate, such as those in the reads skipped by the aligner.

\section{Discussion and Conclusions}

We propose the first lossy reference-free and assembly-free compression approach for FASTQ files, which combines both DNA bases and quality information in the reads to smooth the quality scores and to apply a noise reduction of the bases, while keeping variant calling performance comparable to that with original data.
To the best of our knowledge, there are no tools that have been designed for this purpose, so we compared our results with two
reference-free and read-based tools that only smooth out the quality scores component:
BEETL \cite{JaninRosoneCox2014}
and LEON \cite{LEON2015}.

The resulting FASTQ file with the modified bases and quality scores produced by our tool achieves better compression than the original data. In particular, by using our approach the {\it bases} component achieves better compression than the original (therefore also than competitors which do not make any changes to this component), whereas the compression ratio of the quality scores component is more competitive with BEETL than LEON, which also truncates all quality values greater than ‘@’.
On the other hand, in terms of variant calling, our tool keeps the same accuracy as the original FASTQ data when the ground truth is used as baseline, and preserves the variant calls of the original FASTQ file better than BEETL and LEON.

From the viewpoint of the used resources, LEON has shown to be the fastest tool, although this comparison is not completely fair because our tool modifies different components of the FASTQ file and the outputs are different (not directly comparable).
Moreover, the authors in \cite{LEON2015} state that for WGS datasets, the relative contribution of the Bloom filter is low for high coverage datasets, but prohibitive for low coverage datasets (\eg 10x).
We intend to improve our implementation also using, for instance, parallelization, and test our tool for low coverage datasets and longer reads.

Our implementations give as output the modified FASTQ file, so we have used PPMd and BSC for compression, but we could use any other compressor for this task and we could also combine existing lossless compression schemes to further reduce the size of the FASTQ file, for instance we could use SPRING \cite{SPRING2018}, \fqs \cite{deorowicz2020fqsqueezer}, and others.

Moreover, our strategy could take advantage of the reordering of the reads based on their similarity, \eg as in \cite{CoxBJR12,SPRING2018,deorowicz2020fqsqueezer}. Another feature we did not exploit in our compression scheme is the paired-end information coming from reads in pairs. (Indeed, we compress the FASTQ files in a paired-end dataset independently, as they were single-end.)
Both the above aspects could be analyzed as future work.

We believe the results presented in this paper can motivate the development of new FASTQ compressors that modify the {\it bases} and {\it quality scores} components taking into account both information at the same time to achieve better compression while keeping most of the relevant information in the FASTQ data.
As future work we intend to investigate the error correction problem that needs to 
take into account much more information (\eg reverse-complement, or paired-end information).

\section*{Acknowledgements}

The authors would like to thank N. Prezza for valuable comments and suggestions and for providing part of the code library, and E. Niccoli for preliminary experimental investigations on positional clustering and compression in his bachelor's thesis under the supervision of GR and VG.

Work partially supported by the project MIUR-SIR CMACBioSeq (``Combinatorial Methods for Analysis and Compression of Biological Sequences'') grant n.~RBSI146R5L and by the University of Pisa under the ``PRA – Progetti di Ricerca di Ateneo'' (Institutional Research Grants) - Project no. PRA\_2020\-2021\_26 ``Metodi Informatici Integrati per la Biomedica''.

\end{document}